\documentclass[twocolumn,showpacs,aps,groupedaddress,prd,eqsecnum,epsf]{revtex4}

\usepackage{epsfig}
\usepackage{dcolumn}
\usepackage{bm}
\usepackage{latexsym}
\usepackage{amsfonts}
\usepackage{amssymb}

\newcommand{\beq}{\begin{equation}}
\newcommand{\eeq}{\end{equation}}
\newcommand{\beqa}{\begin{eqnarray}}
\newcommand{\eeqa}{\end{eqnarray}}
\newcommand{\bea}{\begin{array}}
\newcommand{\ena}{\end{array}}

\begin{document}
\draft

\title{Ambiguity of black hole entropy in loop quantum gravity}

\author{Takashi Tamaki$^{1}$}
\email{tamaki@tap.scphys.kyoto-u.ac.jp}
\author{Hidefumi Nomura$^{2}$}
\email{nomura@gravity.phys.waseda.ac.jp}
\address{$^{1}$Department of Physics, Kyoto University, 
606-8501, Japan~\\
$^{2}$Department of Physics, Waseda University, 3-4-1 Okubo,
Shinjuku, Tokyo 169-8555, Japan~ }

\date{\today}

\begin{abstract}
We reexmine some proposals of black hole entropy in loop quantum gravity (LQG) and 
consider a new possible choice of the Immirzi parameter which has not been pointed 
out so far. We also discuss that a new idea is inevitable if we regard the relation 
between the area spectrum in LQG and that in quasinormal mode analysis seriously. 
\end{abstract}

 \maketitle

\section{Introduction}

\begin{figure}[tp]
\psfig{file=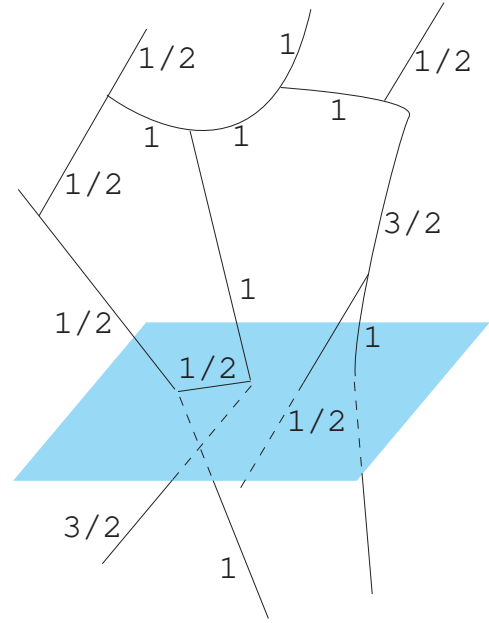,width=2.5in}
\caption{Spin network and a surface.
 \label{SNarea} }
\end{figure}

Loop quantum gravity (LQG) has attracted much attention because of 
its background independent formulation, account for microscopic origin of 
black hole entropy \cite{Corichi}, singularity avoidance in the 
universe \cite{Bojo} and black holes \cite{Bojo2}. 
The spin network has played a key role in the development of this theory \cite{Smolin}. 
Basic ingredients of the spin network are edges. 
In Fig.~\ref{SNarea}, edges are expressed by lines lebeled by 
$j=0$, $1/2$, $1$, $3/2$, $\ldots$ reflecting the SU(2) nature of the gauge group. 
A vertex is an intersection between edges. In this figure, we write only vertices 
where three edges merge (we call them trivalent vertices). 
Even if there is a vertex where more than three edges merge, we can 
decompose it to the sum of edges and trivalent vertices. 
For this reason, we consider only trivalent vertices below. 
For three edges having spin $j_{1}$, $j_{2}$, and $j_{3}$ that merges at an 
arbitrary vertex, we have 
\beqa
\hspace{-10mm}&&j_{1}+j_{2}+j_{3}\in N\ , \label{trivalent}  \\
\hspace{-10mm}&&j_{i}\leq j_{j}+j_{k},\ \   (i,j,k\ \ {\rm different\ from\ each\ other.})
\eeqa
to garantee the gauge invariance of the spin network. 
This is also displayed in Fig.~\ref{SNarea}. 

Using this formalism, general expressions for the spectrum of the area and the 
volume operators can be derived \cite{Rovelli,Ash1}. For example, 
the area spectrum $A$ is 
\beqa
A=4\pi \gamma \sum \sqrt{2j_{i}^{u}(j_{i}^{u}+1)+2j_{i}^{d}(j_{i}^{d}+1)
-j_{i}^{t}(j_{i}^{t}+1)}\ , 
\nonumber
\eeqa
where $\gamma$ is the Immirzi parameter related to an ambiguity 
in the choice of canonically conjugate variables \cite{Immirzi}. 
The sum is added up all intersections between a surface and edges as 
shown in Fig.~\ref{SNarea}. Here, the indices $u$, $d$, and $t$ means edges 
above, below, and tangential to the surface, respectively (We can determine 
which side is above or below arbitrarily). If there is no edges which are 
tangential to the surface, we have $j_{i}^{u}=j_{i}^{d}:=j_{i}$ and 
$j_{i}^{t}=0$. In this case, 
we have the simplified formula as 
\beqa
A=8\pi \gamma \sum \sqrt{j_{i}(j_{i}+1)}\ . 
\label{AreaLQG}
\eeqa
As we will mention below, this is the case for the horizon area spectrum. 

The number of states that determines the black hole entropy is basically 
calculated from (\ref{AreaLQG}) which was first estimated as \cite{Corichi}
\beqa
S= \frac{A\ln (2j_{\rm min}+1)}{8\pi \gamma\sqrt{j_{\rm min}
(j_{\rm min}+1)}}\ ,  \label{SLQG}
\eeqa
where $A$ and $j_{\rm min}$ are the horizon area and the lowest nontrivial representation 
usually taken to be $1/2$ because of SU(2), respectively. In this case, the Immirzi 
parameter is determined as $\gamma =\ln 2/(\pi \sqrt{3})$ to produce the 
Bekenstein-Hawking entropy formula $S=A/4$. 

However, the formula (\ref{SLQG}) was corrected as \cite{Domagala,Meissner}
\beqa
S= \frac{\gamma_{M}A}{4\gamma }\ ,  \label{SLQG2}
\eeqa
where $\gamma_{M}$ is the solution of 
\beqa
1=\sum_{j=Z/2}^{\infty}2\exp (-2\pi\gamma_{M} \sqrt{j(j+1)})\ , \label{Immirzi}
\eeqa
where $j$ takes all the positive half-integer. 
In this case, $\gamma_{M}$ is numerically obtained as $\gamma_{M}=0.23753\cdots$. 
This means that it took four years for the original error to be corrected, 
which suggests that independent reexamining is important. 
Interestingly, other possibilities have also been argued after the result. 
One is to determine $\gamma_{M}$ as the solution of \cite{Alekseev,Khriplovich,Mitra}
\beqa
1=\sum_{j=Z/2}^{\infty}(2j+1)\exp (-2\pi\gamma_{M} \sqrt{j(j+1)})\ . 
\label{Immirzi2}
\eeqa
In this case, $\gamma_{M}=0.27398\cdots$.
The other is to recover (\ref{SLQG}) by imposing the condition that the 
area is constructed only by $j=j_{\rm min}$~\cite{Dreyer2}. 

These provide us with the following question: that is, which is the best 
choice for the Immirzi parameter? Therefore, we reanalyze these possibilities.  
This is important in the following reasons. 

(i) In string theory, number counting for microscopic states of black holes 
has been considered, and it has reproduced the Bekenstein-Hawking formula $S=A/4$ 
\cite{Strominger}.
In the future, it is desirable for us to have a connection with the number counting in 
string theory. Although there is no relation between LQG and string theory at 
present, this may shed new light on the developments to come in theoretical physics. 
Probably, we will need to proceed many steps toward this purpose. 

However, there is a subject which can be attacked soon. This is 
(ii) the possible relation to the quasinormal mode which has been argued as 
another consistency check of the Immirzi parameter in the area spectrum \cite{Schiappa}. 
Using (\ref{SLQG}), an encounter between LQG and the quasinormal mode was 
considered first in Ref.~\cite{Dreyer}. This means that 
if we have $j_{\rm min}=1$, we can determine $\gamma$ as $\ln 3/(2\pi\sqrt{2})$ 
which gives $A=4\ln 3$ from (\ref{AreaLQG}). 
This coincides the area spectrum determined by quasinormal mode using 
Bohr's correspondence \cite{Hod}. 
Moreover, the quasinormal mode analysis that originally performed in 
Schwarzschild black hole \cite{Motl,Kuns,Birm} has been extended to 
single-horizon black holes \cite{Tamaki,Visser,Pad,Kettner,Das}. 
These results suggest that there {\it is} a relation between these spectra. 
However, if we adopt (\ref{Immirzi}) or (\ref{Immirzi2}), we cannot obtain 
such a consistency. Thus, we also want to know which is the best choice for 
the Immirzi parameter in this view point. 

This paper is organized as follows. In Sec. II, we summarize the framework 
\cite{Corichi} (which we call the ABCK framework.) that 
is necessary in number counting. In Sec. III, we argue various possibilities 
that determine the number of state. In Sec. IV, we summarize our results and 
discuss their meaning. 

\section{Summary of the ABCK framework}

Here, we briefly introduce the framework in Ref.~\cite{Corichi} and 
summarize the conditions necessary for number counting. 
First, we introduce the isolated horizon (IH) where we can reduce the SU(2) 
connection to the U(1) connection. This plays the important role of determining 
the conditions (ii) and (iii) below. For details, see \cite{isolated}. 
Moreover, the merit of the IH is that we can treat the event horizon 
and the cosmological horizon, where we can define the Hawking temperature 
in an unified way. 

Next, we imagine that spin network pierces the IH. By eliminating the edge 
tangential to the isolated horizon, we can decompose the Hilbert space as 
the tensor product of that in the IH $H_{IH}$ and that in the 
bulk $H_{\Sigma}$, i.e., $H_{IH}\otimes H_{\Sigma}$. If we specify the points 
that are intersections of edges having spin $(j_{1},j_{2},\cdots ,j_{n})$ 
and the IH, we can write $H_{\Sigma}$ as the orthogonal sum 
\begin{equation}
H_{\Sigma}=\bigoplus_{j_{i},m_{i}}H_{\Sigma}^{j_{i},m_{i}}\ , \label{sum}
\end{equation}
where $m_{i}$ takes the value $-j_{i}$, $-j_{i}+1$, $\cdots$, $j_{i}$. 
This is related to the flux operator eigenvalue $e_{s'}^{m_{i}}$ 
that is normal to the IH ($s'$ is the part of the IH that have 
only one intersection between the edge with spin $j_{i}$.)
\begin{equation}
e_{s'}^{m_{i}}=8\pi\gamma m_{i}\ . \label{flux}
\end{equation}
Since we eliminate the edge tangential to the IH, we have $m_{i}\neq 0$. 
That is also the reason why the area spectrum 
is simplified as (\ref{AreaLQG}). The horizon Hilbert space can be written 
as the orthogonal sum similar to (\ref{sum}) by eigenstates 
$\Psi_{b}$ of the holonomy operator $\hat{h}_{i}$, i.e., 
\beqa
\hat{h}_{i}\Psi_{b}=e^{\frac{2\pi ib_{i}}{k}}\Psi_{b}\ . 
\label{holonomy}
\eeqa

Next, we consider the constraints in the bulk and at the IH, respectively. 
In the bulk, the Gauss constraint is already satisfied and the diffeomorphism 
constraint means that the place to which the edges stick the IH is not relavant. 
The scalar constraint is non-trivial. However, 
since $(j,m)$ characterize the bulk almost at the IH, it is assumed that 
the bulk scalar constraint does not affect $(j,m)$. 
At the IH, we do not consider the scalar constraint since the lapse function 
disappears. If we require that the horizon should be invariant under the 
diffeomorphism and the U(1) gauge transformation, 
The horizon area $A$ is fixed as 
\begin{equation}
A=4\pi \gamma k\ \ , \label{area}
\end{equation}
where $k$ is natural number and it is the level of the Chern-Simons theory. 
In addition to this condition, it is required that we should fix 
an ordering $(j_{1},j_{2},\cdots ,j_{n})$. 
The area operator eigenvalue $A_{j}$ should satisfy 
\beqa
(i)\ \ A_{j}=8\pi \gamma \displaystyle  \sum_{i} \sqrt{j_i(j_i+1)}\leq A\ . 
\label{AreaLQG2}
\eeqa

We mention other conditions. From the quantum Gauss-Bonnet theorem, we require 
\beqa
(ii)\ \ \sum_{i=1}^{n}b_{i}=0\ . 
\label{GB}
\eeqa
From the boundary condition between the IH and the bulk, we have 
\beqa
(iii)\ \ b_{i}=-2m_{i}\ \ \ {\rm mod} k\ . 
\label{boundary}
\eeqa
All we need to consider in number counting are (i)(ii)(iii). 

\section{number counting}

Here, we consider number counting based on the ABCK framework. 
If we use (ii) and (iii), we obtain 
\beqa
(ii)'\ \ \sum_{i=1}^{n}m_{i}=n'\frac{k}{2}\ . 
\label{GB2}
\eeqa
In \cite{Meissner}, it was shown that this condition is irrelevant in number 
counting. Thus, we perform number counting only concentrating on (i) below. 

For this purpose, there are two different points of view. 
The one adopted in the original paper \cite{Corichi,Domagala,Meissner} 
counts the {\it surface} freedom 
$(b_{1},b_{2},\cdots ,b_{n})$. The second 
counts the freedom for both $j$ and $b$ \cite{Khriplovich,Mitra}. 

We first consider the second possibility since (we suppose) it is easier to 
understand.  To simplify the problem, we first consider the set $M_{k}$ by 
following \cite{Domagala}, that is 
\beqa
M_{k}:=\left\{(j_{1},\cdots, j_{n})|0\neq j_{i}\in \frac{Z}{2}, 
\displaystyle  \sum_{i} j_i \leq \frac{k}{2}
\right\}\ . 
\label{MK}
\eeqa
Here, we also eliminate $A$ using (\ref{area}). Let $N_{k}$ be 
the number of elements of $M_{k}$ plus $1$. Certainly, 
\beqa
N(a)\leq N_{k}\ , \label{upper}
\eeqa
where $N(a)$ ($a:=\frac{A}{8\pi\gamma}$) is the number of states which account 
for the entropy. Note that if $(j_{1},\cdots,j_{n})\in M_{k-1}$, then  
$(j_{1},\cdots,j_{n},\frac{1}{2})\in M_{k}$. 
In the same way, for natural $0<s\leq k$, 
\beqa
(j_{1},\cdots,j_{n})\in M_{k-s}\Rightarrow (j_{1},\cdots,j_{n},
\frac{s}{2})\in M_{k}\ . \label{relation1}
\eeqa
Then, if we consider all $0<s\leq k$ and all the sequence 
$(j_{1},\cdots,j_{n})\in M_{k-s}$, we found that $(j_{1},\cdots,j_{n}, 
\frac{s}{2})$ form the entire set $M_{k}$. Moreover, for $s\neq s'$, 
\beqa
(j_{1},\cdots,j_{n},\frac{s}{2})\neq (j_{1},\cdots,j_{n},
\frac{s'}{2})\in M_{k}\ . \label{relation2}
\eeqa
The important point to remember is that we should include 
the condition $m_{i}\neq 0$ (or equibalently $b_{i}\neq 0$). Thus, 
each $j_{i}$ has freedom $2j_{i}$ for the $j_{i}$ integer and the $2j_{i}+1$ way 
for the $j_{i}$ half-integer. They are 
summarized as $2[\frac{2j+1}{2}]$ where $[\cdots ]$ is the integer parts. 
For this reason, the recursion relation is
\beqa
N_{k}=\sum_{s=1}2[\frac{s+1}{2}](N_{k-s}-1)+1\ . \label{recursion}
\eeqa
This is the point which has not been examined out so far.

As a stright forward extension of this, we can consider $N(a)$, which is  
\beqa
\hspace{-5mm}N(a):=\left\{(j_{1},\cdots, j_{n})|0\neq j_{i}\in \frac{Z}{2}, \right. 
\nonumber \\
\hspace{-5mm}\left.\displaystyle  \sum_{i} \sqrt{j_i (j_i +1)}\leq 
\frac{k}{2}=a
\right\}\ . 
\label{number2}
\eeqa
In this case, we obtain the recursion relation 
\beqa
\hspace{-10mm}&&N (a)=2N(a-\sqrt{3}/2)+2N(a-\sqrt{2})+\cdots + 
\nonumber  \\
\hspace{-10mm}&&2[\frac{2j+1}{2}]N(a-\sqrt{j_i (j_i +1)})
+\cdots +[\sqrt{4a^{2}+1}-1]\ . \label{recursion2}
\eeqa
If we notice that the solution of $\sqrt{j_i (j_i +1)}=a$ is 
$j_i =(\sqrt{4a^{2}+1}-1)/2$, 
meaning of $[\sqrt{4a^{2}+1}-1]$ is obvious. 

If we use the relation
\beqa
N(a)=Ce^{ \frac{A\gamma_{M}}{4\gamma} }\ , \label{number-behavior}
\eeqa
where $C$ is a constant, that was obtained in \cite{Meissner}, we obtain 
\beqa
1=\sum_{j=Z/2}2[\frac{2j+1}{2}]\exp (-2\pi\gamma_{M} \sqrt{j(j+1)})\ ,
\label{Immirzi3}
\eeqa
by plugging (\ref{number-behavior}) into (\ref{recursion2}) and 
taking the limit $A\to\infty$. Then if we require $S=A/4$, we have 
$\gamma =\gamma_{M}$. This is the extension of \cite{Mitra,Khriplovich}. 
In this case, $\gamma =0.26196\cdots$. 

Next, we consider the first possibility that counts only the surface freedom. 
This means that even if $(j_{1},j_{2},\cdots ,j_{n})$ is different, 
it is regareded as the {\it same} surface state if the horizon area and 
$(b_{1},b_{2},\cdots ,b_{n})$ are same. For example, $(j_{1},j_{2})=(3/2,1/2)$ 
and $(1/2,3/2)$ both give the possibility $(b_{1},b_{2})=(-1,-1)$. 
Then, it should {\it not} be distinguished in this description. 

What should we do in this number counting ? This following is the method taken in 
\cite{Domagala}, i.e., we rewrite (\ref{AreaLQG2}) as 
\beqa
(i)'\ \ 8\pi \gamma \displaystyle  \sum_{i} \sqrt{|m_i |(|m_i |+1)}
\leq A\ . 
\label{AreaLQG3}
\eeqa
Let us compare (\ref{relation1}) with 
\beqa
\hspace{-5mm}(m_{1},\cdots,m_{n})\in M_{k-s}\Rightarrow (m_{1},\cdots,m_{n},
\pm \frac{s}{2})\in M_{k}\ . \label{relation1DL}
\eeqa
At first glance, it might seem that we abondon the freedom 
$m_{n+1}=-\frac{s}{2}+1,\cdots \frac{s}{2}-1$. However, it is not the case 
since we obtain that freedom from $M_{k-s+2}$, $M_{k-s+4}$, $\cdots$. 
It is the crucial difference from (\ref{relation1}) where the freedom of $j$ 
is counted. In this way, we have the relation 
\beqa
N_{k}=\sum_{s=1}2(N_{k-s}-1)+1\ .   \label{recursion-DL}
\eeqa
Therefore, we obtain (\ref{Immirzi}).

\section{conclusions and discussion}

In this paper, we have considered two possibilities for the 
number of states of black holes in the ABCK framework. One of them gives a new value for 
the Immirzi parameter. From these results, we consider whether or not 
there is a consistency between the area spectrum in LQG and the area spectrum in the 
quasinormal mode. Since the area spectrum obtained 
from the quasinormal mode is $dA=4\ln 3$, it is obvious that we do not 
have the same consistency if we adopt the Immirzi parameter determined by (\ref{Immirzi}) 
or (\ref{Immirzi3}). Then, how about the case in which only $j=j_{\rm min}$ 
survives, as considered in \cite{Dreyer2} ? Unfortunately, both 
(\ref{Immirzi}) and (\ref{Immirzi3}) do not provide 
consistency that is different from the case in (\ref{Immirzi2}). 
This means that if we take the consistency to the 
quasinormal mode seriously, we will need new considerations. 

Finally, we want to consider which of the 
two candidates is the better choice. The reason why only surface degree was counted in 
\cite{Corichi,Domagala,Meissner} is to separate surface degree from the bulk 
freedom. If we admit $j$ as an independent variable, it is difficult to 
separate it from other bulk freedoms since that in the bulk can communicate 
with infinity. However, as pointed out in \cite{Mitra}, it is $j$ 
that determines area eigenvalue and other bulk variables are irrelevant. 
Moreover, since 
quantum horizons would fluctuate \cite{Rafal}, it may be a problem 
to consider the IH as a sharp boundary. For these reasons, 
it is too early to abondon the possibility that we could count $j$ as an independent 
variable. Of course, it is also important to consider the other method 
in the calculating the number of freedom as in \cite{Dasgupta}. 
We also want to examine these possibilities in future.

\acknowledgements
We would like to thank Professor Jerzy Lewandowski for useful comments 
and thank Professor Lee Smolin for hospitality when one of us stayed 
in Perimeter Institute. 
This work was partially supported by The 21st Century COE
Program (Holistic Research and Education Center for 
Physics Self-Organization Systems) at Waseda University.
This work was supported in part by a JSPS Grant-in-Aid, No.\ 154568(T.T.).  
This work was also supported in part by a
Grant-in-Aid for the 21st Century COE ``Center for Diversity and
Universality in Physics".

%


\begin{references}
\bibitem{Corichi}
A. Ashtekar, J. Baez, A. Corichi, and K. Krasnov, 
Phys. Rev. Lett. {\bf 80}, 904 (1998); 
A. Ashtekar, J. Baez, and K. Krasnov, Adv. Theor. Math. 
Phys. {\bf 4}, 1 (2000). 
\bibitem{Bojo}
For review, see, e.g., M. Bojowald and H. A. Morales-Tecotl, 
Lect. Notes Phys. {\bf 646}, 421 (2004).
\bibitem{Bojo2}
M. Bojowald, Phys. Rev. Lett. {\bf 95}, 61301 (2005). 
\bibitem{Smolin}
C. Rovelli and L. Smolin, Phys. Rev. D {\bf 52}, 5743 (1995). 
\bibitem{Rovelli}
C. Rovelli and L. Smolin, Nucl. Phys. B {\bf 442}, 593 (1995); 
Erratum, {\it ibid.}, {\bf 456}, 753 (1995). 
\bibitem{Ash1}
A. Ashtekar and J. Lewandowski, Class. Quantum Grav. {\bf 14}, 
A55 (1997). 
\bibitem{Immirzi}
G. Immirzi, Nucl. Phys. Proc. Suppl. B {\bf 57}, 65 (1997). 
\bibitem{Domagala}
M. Domagala and J. Lewandowski,
Class. Quant. Grav. {\bf 21}, 5233 (2004). 
\bibitem{Meissner}
K. A. Meissner,
Class. Quant. Grav. {\bf 21}, 5245 (2004). 
\bibitem{Alekseev}
A. Alekseev, A. P. Polychronakos, and M. Smedback, Phys. Lett. B 
{\bf 574}, 296 (2003); A. P. Polychronakos, Phys. Rev. D {\bf 69}, 
044010 (2004). 
\bibitem{Khriplovich}
I.B. Khriplovich, gr-qc/0409031;
gr-qc/0411109. 
\bibitem{Mitra}
A. Ghosh and P. Mitra, Phys. Lett. B {\bf 616}, 114 (2005). 
\bibitem{Dreyer2}
S. Alexandrov, gr-qc/0408033 (2004); 
O. Dreyer, F. Markopoulou and L. Smolin
hep-th/0409056 (2004).
\bibitem{Strominger}
A. Strominger and C. Vafa, Phys. Lett. B {\bf 379}, 99 (1996); 
J. M. Maldacena and A. Strominger, Phys. Rev. Lett. {\bf 77}, 428 (1996). 
\bibitem{Schiappa}
For review, see, e.g., J. Natario and R. Schiappa,
hep-th/0411267.
\bibitem{Dreyer}
O. Dreyer, Phys. Rev. Lett. {\bf 90}, 081301 (2003). 
\bibitem{Hod}
S. Hod, Phys. Rev. Lett. {\bf 81}, 4293 (1998). 
\bibitem{Motl}
L. Motl, Adv. Theor. Math. Phys. {\bf 6}, 1135 (2003); 
L. Motl and A. Neitzke, {\it ibid.}, {\bf 7}, 307 (2003). 
\bibitem{Kuns}
G. Kunstatter, Phys. Rev. Lett. {\bf 90}, 161301 (2003). 
\bibitem{Birm}
D. Birmingham, Phys. Lett. B {\bf 569}, 199 (2003). 
\bibitem{Tamaki}
T. Tamaki and H. Nomura,
Phys. Rev. D {\bf 70}, 044041 (2004). 
\bibitem{Visser}
A. J. M. Medved, D. Martin, and M. Visser, Class. Quantum Grav. {\bf 21}, 
1393 (2004); {\it ibid.}, 2393 (2004). 
\bibitem{Pad}
T. Padmanabhan, Class. Quantum Grav. {\bf 21}, L1 (2004); 
T. R. Choudhury and T. Padmanabhan, Phys. Rev. D {\bf 69}, 064033 (2004). 
\bibitem{Kettner}
J. Kettner, G. Kunstatter, and A.J.M. Medved,
Class. Quant. Grav. {\bf 21}, 5317 (2004).
\bibitem{Das}
S. Das and S. Shankaranarayanan,
Class. Quant. Grav. {\bf 22} (2005) L7.
\bibitem{isolated}
A. Ashtekar, A. Corichi, and K. Krasnov, Adv. Theor. Math. 
Phys. {\bf 3}, 419 (1999). 
\bibitem{Rafal}
M. Bojowald and R. Swiderski, Phys. Rev. D {\bf 71}, 081501 (2005). 
\bibitem{Dasgupta}
A. Dasgupta, JCAP08, 004 (2003).
\end{references}
\end{document}